# Energy-resolved hot carrier relaxation dynamics in monocrystalline plasmonic nanoantennas

Régis Méjard, Anthonin Verdy, Marlène Petit, Alexandre Bouhelier, Benoît Cluzel and Olivier Demichel

*Laboratoire Interdisciplinaire Carnot de Bourgogne, UMR 6303 CNRS-Université Bourgogne Franche-Comté, 21078 Dijon, France*

Corresponding Author
Email: olivier.demichel@u-bourgogne.fr

**ABSTRACT**
Hot carriers are energetic photo-excited carriers driving a large range of chemico-physical mechanisms. At the nanoscale, an efficient generation of these carriers is facilitated by illuminating plasmonic antennas. However, the ultrafast relaxation rate severally impedes their deployment in future hot-carrier based devices. In this paper, we report on the picosecond relaxation dynamics of hot carriers in plasmonic monocrystalline gold nanoantennas. The temporal dynamics of the hot carriers is experimentally investigated by interrogating the nonlinear photoluminescence response of the antenna with a spectrally-resolved two-pulse correlation configuration. We measure time-dependent nonlinearity orders varying from 1 to 8, which challenge the common interpretation of multi-photon gold luminescence. We demonstrate that the relaxation of the photo-excited carriers depends of their energies relative to the Fermi level. We find a 60 % variation in the relaxation rate for electron-hole pair energies ranging from c.a. 0.2 to 1.8 eV. The quantitative relationship between hot carrier energy and relaxation dynamics is an important finding for optimizing hot carriers-assisted processes and shed new light in the intricacy of nonlinear photoluminescence in plasmonic structures.

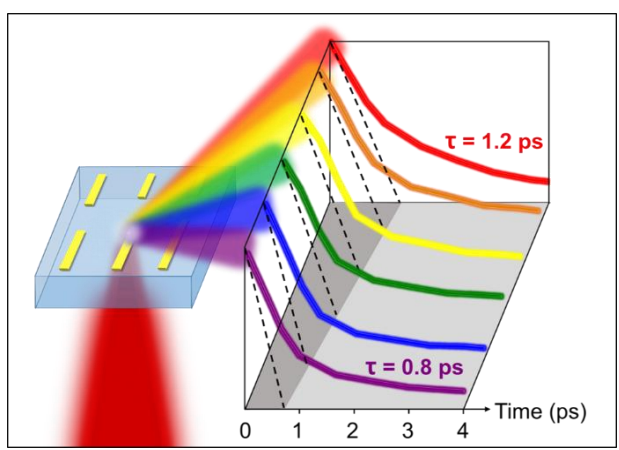

**KEY WORDS:** *Autocorrelation, ultrafast dynamics, nonlinear plasmonics, monocrystalline gold, nanoantennas, hot carriers, photoluminescence*

When a light pulse illuminates a metal, photons may excite electrons to energy levels well above the Fermi level. These out-of-equilibrium electron-hole pairs, referred to as hot carriers, carry enough energy to potentially drive chemico-physical processes with applications ranging from drug delivery to nanowire growth and chemical catalysis.[1] Photons may also couple to surface plasmons, which are collective oscillations of the electronic cloud. The excitation of localized surface plasmons in metal nanoantennas enhance by many-fold the interaction cross-sections with incoming photons.[2,3] Thus, the large electromagnetic fields localized at the nanoantenna contribute to the creation of an energetic non-thermal Fermi-Dirac distribution, which depends on the material and the geometry of the antenna.[1] A significant research effort currently focuses on the production and extraction of these non-thermal carriers as reviewed in Refs[1,4-6]. The temporal evolution of the hot-carrier distribution follows a well-established dynamics. After absorbing a photon, the photo-excited carriers are swiftly thermalized through carrier-carrier scattering with a characteristic rate $\tau_{th}$ lasting a few hundreds of femtoseconds. After internal thermalization, the hot carriers couple their energy to the phonon population and the carrier distribution decays with a rate $\tau_{e-ph}$ with a timescale of a few picoseconds.[7,8] Spatial and temporal aspects play an essential role in the harvest of hot carriers, thus the knowledge and management of the hot carrier dynamics is fundamental to envisage the future development of efficient plasmon-based hot carrier devices. The excitation of plasmonic resonances not only favours the efficiency of generating hot carriers, but also triggers a variety of nonlinear optical responses from the material itself.[9] In the context of the present discussion, the multi-photon photo-luminescence (MPL) emitted from gold optical antennas is particularly interesting. The exact mechanisms giving rise to the nonlinear luminescence is still debated. In one view, the nonlinearity may arise following the absorption of multiple photons[10-17] and the transition sequence involves a real intermediate state as well as a final state populated by hot carriers[18,8]. Another view consists in explaining the nonlinearity by the thermodynamics of the intraband relaxation of a hot electron distribution. The nonlinearity stems from the emission process rather than multi-photon absorption events.[19] Regardless of the exact understanding of MPL, the generation efficiency of the hot carriers and their dynamics are related to the nonlinear luminescence intensity and temporal response.[7]

In this article, we report on spectrally resolved MPL auto-correlation measurements conducted on single-crystal gold nanoantennas. Although electron-electron scattering is an important process for extracting non-thermal carriers[4-6], we restrict our analysis to carrier-phonon scattering rates and we extract a hot-carrier relaxation time in the picosecond range. Interestingly, we found that the hot carrier dynamics is 60 % larger for carriers with energies close to the Fermi level ($E_f$) than for energies 1.5 eV above. The observation of an energy-dependent kinetics is an important parameter for controlling the energy transferred from a hot carrier distribution.

**EXPERIMENTAL DETAILS**

With the final objective to provide an optimal device, the nature and the quality of the materials employed for plasmonic nanostructures is first considered. To date, plasmonics has mainly relied on the deposition of polycrystalline gold thin films (c.a. 50 nm) onto a thin adhesion promoter layer (chromium, titanium) under high vacuum. However, Mc Peack *et al.* recently reported that this approach, although convenient on practical regards, leads to significant discrepancies between samples[20], which could become even more important when comparing samples from groups around the world. In particular, the conditions of vacuum as well as the deposition rate and the purity of the target generate various surface roughness and grain sizes in polycrystalline gold. These parameters increase the damping of plasmons thus broadening plasmonic resonance[21] and reducing the plasmon propagation length[22]. These parameters have also been identified as altering the electron-phonon relaxation rate.[23] Furthermore, because it is ultrathin, the role played by the adhesion promoter is often neglected although some metal interdiffusion may change the overall optical properties of the structures over time.[24] Finally, when plasmonic designs involve substrates with low wettability, the topography of the gold layer, and thus its permittivity, can drastically depart from predictions.[25] To sum up, standard gold deposition approaches are detrimental to the reliability of potential plasmonic technologies due to their non-reproducibility. On the contrary, structures obtained by colloidal synthesis are single crystals[26-28] and therefore are identical from one synthesis to another independently of the equipment and the personnel. In this work, we employ micron-size monocrystalline plates as a base material to create the desired nanostructures in a top-down approach.[29-31] This ensures reproducible investigations on nanostructures that are known to be more accurate in terms of shape and topography aspects.[29,30,32] The present work deals with elementary structures with the aspiration to pave the way to build up robust, more complex systems.

We, first, present our study on a large monocrystalline plate and the nonlinear photoluminescence it generates. We subsequently move on to spectrally resolved autocorrelation measurements obtained on a monocrystalline nanoantenna.

We synthesize monocrystalline gold plates from which we realize an array of plasmonic nanorods. Figure 1.a shows a transmission electron microscopy (TEM) image of a typical micron-size Au plate produced by the synthesis. The colloidal synthesis is based on the generic reaction reducing the oxidized form of gold $Au^{3+}(Cl^-)_4$.[33] We have adapted the procedure of Ref.[27] to produce plates of a few tens of nanometres in thickness (30-80 nm) and of few tens of micrometres in the lateral directions. Briefly, a solution of ethylene glycol is brought to 65 °C. A solution of 0.036 mmol of $HauCl_4$ diluted in ethylene glycol and then 0.072 mmol of aniline are then introduced in the vessel. The reduction of gold into plates is completed in 3 to 4 hours. High resolution TEM confirms the monocrystallinity of the plates and the {111} orientation of the top facets. Additionally, energy dispersive X-ray (EDX) analyses validate that the plates are composed of gold with no other chemical compounds from the synthesis. The analyses ensure the high quality of the material and its universal nature.

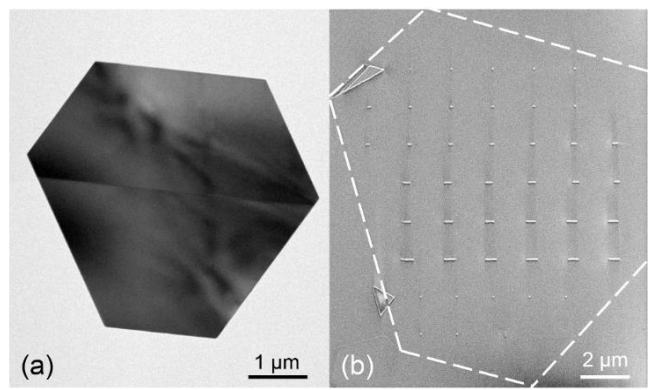

Figure 1: (a) Transmission electron microscopy (TEM) image of a single-crystal gold nanoplate, (b) Scanning electron microscopy image of an array of single-crystal nanoantennas etched in a single-crystal gold plate on a glass substrate. The dashed line represents the boundaries of the plate before etching.

After the colloidal synthesis, the monocrystals are dispersed in an ethanol solution and dropcasted on a glass substrate. Gold plates of about 50 nm in thickness are selected by atomic force microscopy measurements. Next, a standard electron-beam lithography with an adapted alignment procedure is conducted to define the layout of optical antennas on the selected plate. A Cr layer is then evaporated by vapour deposition to create a mask prior to removing the electron-beam resist. The antennas are then finalized by reactive ion etching of the gold plate and a removal of the Cr layer. Further detail of this procedure will be reported somewhere else. Figure 1.b. illustrates an array of gold nanorods with lengths varying from 90 to 615 nm and width fixed at 85 ± 7 nm (standard error) etched from a single crystal gold plate.

Figure 2 sketches the optical setup employed in this work. The nonlinear optical processes emitted from monocrystalline gold plates and antennas are produced upon irradiation of a near-IR (820 nm – 1.51 eV) pulsed Ti:Sapphire laser with a pulse length of 120 fs. The laser pulses are focused through a high numerical aperture oil-immersion objective (NA = 1.49, 60x) to produce a diffraction limited excitation spot of about 300 nm in diameter enabling the investigation of a unique nanostructure. The nonlinear emission is collected by an optical fibre in a plane conjugate to the focal plane which acts as a spatial filter to capture only the light produced around the excitation region. The dispersion introduced by the optics, in particular due to the objective, is precompensated by a 4-f zero dispersion line[34] to ensure Fourier-transform limited, chirp-free optical pulses in the objective's focal plane. Auto-correlation measurements are performed with a homemade Michelson interferometer producing two delayed pulses. The intensity of the pulses may be varied independently. The nonlinear signal is recorded as a function of the delay on a spectrometer after filtering out the laser line.

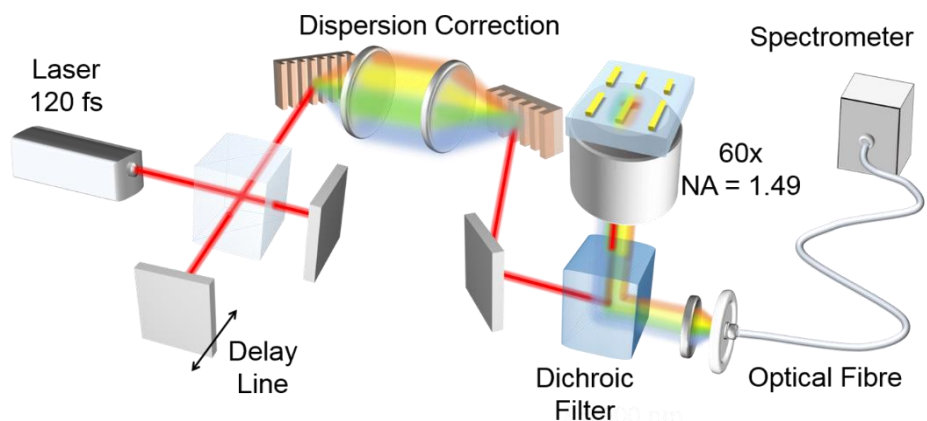

Figure 2: Schematic of the optical setup. A delay line consisting of a Michelson interferometer outputs two delayed pulses of c.a. 120 fs at 820 nm. The pulses are chirp-compensated with a 4-f zero dispersion line. The pulses are then sent to the sample through a high-NA objective. The nonlinear signal emitted by the sample is filtered and directed to an optical fibre before reaching the spectrometer.

**RESULTS AND DISCUSSION**

First, we explore the nonlinear optical properties of a large gold plate for which no localized surface plasmon resonance occurs at the excitation wavelength due to its micrometre size. We record hyperspectral nonlinear images, which consist at recording the nonlinear spectrum emitted by the object for each excitation position.[35] Gold produces two main nonlinear processes: second harmonic generation (SHG) and MPL. SHG is a coherent, instantaneous mechanism. It possesses an optical spectrum as narrow as the pulse spectrum divided by √2. MPL is an incoherent mechanism featuring a wide spectrum spanning over the whole spectral detection window of our spectrometer (375 – 725 nm). Figure 3.a show confocal images of the nonlinear responses integrated over the spectral ranges of SHG and MPL. Note that two areas are saturated in Figure 3.a. The one located at the bottom left of the plate is where the adjustment of the laser focalisation was performed. The structure is likely to be locally damaged by the long laser exposure.[36] The second area, located in the middle of the right edge is a response of nanoparticle adsorbed on surface. Both areas are irrelevant to our study and are disregarded when setting the best intensity scale for visualisation clarity. The SHG response is somewhat intense in the central region, although the crystal is purely centrosymmetric and the theoretically expected SHG signal is zero for symmetry reasons.[37] This is an indication that the origin of the SHG in monocrystalline plates may be related to the centrosymmetry breakup at the interface.[38] Another interpretation could be that electromagnetic field gradients play a non-negligible part in this mechanism.[39]

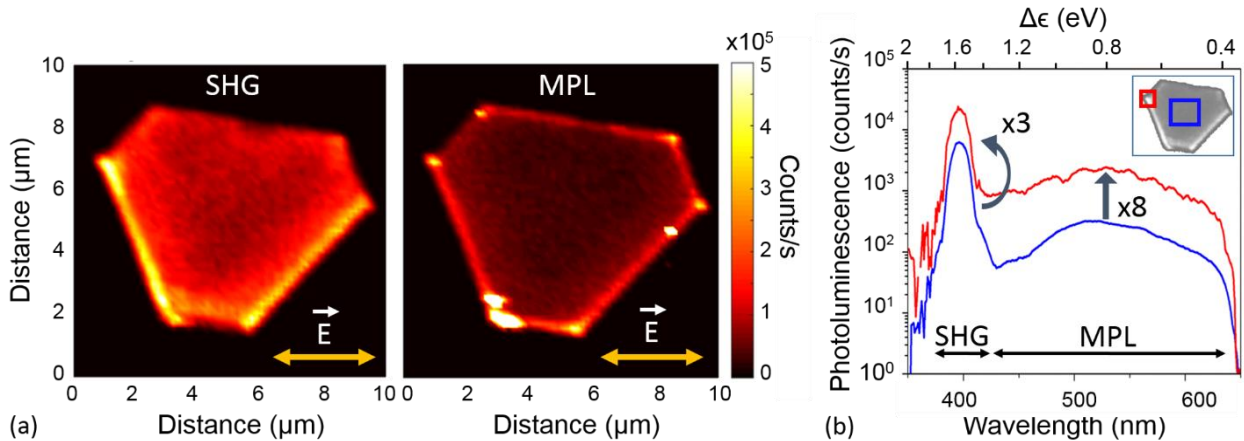

Figure 3: Two confocal maps of a single-crystal Au plate. a) SHG- and MPL-filtered images. The double arrow represents the polarization. b) Spectra measured at the centre and a corner of the plate (the inset shows the two locations). The top horizontal axis indicates the lower boundary of the electron-hole pair energy: $\Delta\epsilon = hc \cdot (1/\lambda - 1/\lambda_{laser})$.

The response from the edges differs depending on the nonlinear process: MPL is enhanced at every edge regardless of its orientation and is strongest at the structure corners. The behaviour is much different for SHG which is mainly enhanced for a few edges. The SH bright edges are depending on several parameters, namely the orientation of the crystal lattice, the polarization angle[40] and the propagation of running surface plasmons[41]. The MPL image appears more contrasted than the SHG image as confirmed by the spectra of Figure 3.b. The MPL spectrum features an eight-fold amplitude rise from the plate's central region to a corner position whereas the SHG signal only increases by a factor three. The highest sensitivity of the MPL compared to the SHG to the field variations at the corners suggests a higher nonlinearity order as already observed.[8,14,15] Note that in Figure 3.b, the energy is calculated as follows: $\Delta\epsilon = hc \cdot (1/\lambda - 1/\lambda_{laser})$, where $\lambda$ is the wavelength of the nonlinear emission and $\lambda_{laser}$ is the wavelength of the laser. $\Delta\epsilon$ represents the lower boundary of the energy gap of electron-hole pairs recombining radiatively.

In order to determine the nonlinearity order of both mechanisms, the SHG and MPL intensities are recorded as a function of the laser power. The power dependences are reported in figure 4.a. In logarithmic scales, both mechanisms follow straight lines with slopes indicating the nonlinearity order. As expected, the SHG nonlinearity order is 2 whereas that of MPL was measured to be close to 3. Such high power-law exponents have already been reported previously for MPL and is a consequence of very short excitation pulses.[18] Biagioni *et al.* demonstrated nonlinearity orders varying from 2 to 4 when the pulse duration was shortened from about 700 fs to 40 fs.[8]

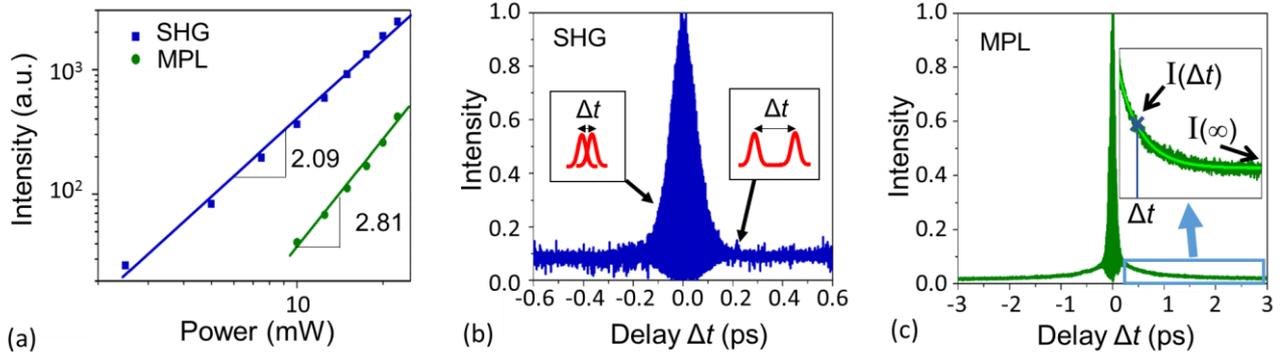

Figure 4: Nonlinear signals from single-crystal plates. (a) Power dependence of SHG and MPL intensity, (b) SHG and (c) MPL autocorrelation measurements. Inset of (c) is a magnification of the right decay wing with a monoexponential decay fit overlayed.

We now turn our attention to the core of this letter by discussing the hot carrier-lattice relaxation dynamics. We set the power of the two pulses produced by the Michelson interferometer to be equal. Figure 4.b (resp. c) shows an auto-correlation measurement obtained by spectrally selecting the SHG (resp. MPL) contribution. SHG is an instantaneous and coherent process and, as such, SHG auto-correlation traces give a direct information on the temporal shape of the pulse in the exact focal plane. Figure 4.b shows a typical autocorrelation trace for unchirped pulses of 120 fs. This is the confirmation of an adequate chirp compensation and therefore of Fourier-transform-limited pulses in the sample plane. This technical point is particularly important to calibrate the setup and avoid any artefact in further quantitative analysis of the MPL signal.

In a typical case of a single-pulse excitation, MPL is an incoherent process that involves a hot carrier distribution.[7,8,19,42] The hot carriers created sustain carrier-carrier scattering during the first hundreds of fs[43-45], which is not accessible with our setup and vanishes before ps timescales. The hot-carrier bath can therefore be considered in thermal equilibrium, described by a Fermi-Dirac distribution at a temperature $T_e$. Some of the hot carrier population, then scatter on the lattice and create phonons for a non-zero characteristic period of time, $\tau_{e-ph}$. The lattice temperature is assumed to be at room temperature, $T_0$, as the lattice relaxation time is very slow compare to the ps timescale. The standard two-temperature model describes the time-dependent evolution of the electronic temperature.[46]

In the autocorrelation measurement, a second pulse illuminates the structure and produces its own MPL emission independently of the first one if the interpulse delay, $\Delta t$, is large ($\Delta t$ ~10 ps). However, if the second pulse illuminates the structure with $\Delta t < \tau$, additional hot carriers are excited by the delayed second pulse to produce additional MPL. The nature of the MPL involving a heated bath of photoexcited carriers of finite lifetime is the hallmark of the wings appearing in the autocorrelation trace for delays $\Delta t > 300$ fs where the two pulses are temporally separated. These decaying wings are clearly visible in our MPL measurements as reported in Figure 4.c in stark contrast with the SHG measurement in figure 4.b.[7] According to the classical two-temperature model[46] the MPL intensity of the decaying wings, $\Delta I = I(\Delta t) - I(\infty)$, can be fitted with a monoexponential as a function of the interpulse delay $\Delta t$.[18] Hence, in this description, autocorrelation gives a direct quantitative measure of the hot carrier relaxation dynamics.[7,8] Our measurements are typically very well fitted with exponential decay curves ($R^2$ ~ 0.99) and the characteristic decay times lie in the picosecond range. Therefore our measurements are consistent with hot carrier relaxation through carrier-phonon interactions in gold[47,48] and validate the aforementioned assumption underlying the two-temperature model.

Based on this analysis, we now turn to optical antennas etched from a single-crystal plate. The nanoantennas exhibit stronger MPL emission indicating a better hot carrier generation efficiency compared to the planar structure.[18] By comparing the photoluminescence intensity obtained from a gold plate and from an out-of-resonance nanoantenna with the same excitation, we evaluate an efficiency enhancement on the order of $10^3$. Auto-correlation measurements on gold nanostructures have already been reported with characteristic timescale in the picosecond range.[7,8] However, to date, spectrally resolved autocorrelation measurements, referred to as coloured autocorrelations, have not been reported yet despite their potential to improve the understanding of hot carrier dynamics at play in plasmonic nanostructures. We study a nanoantenna of 373 nm in length, 81 nm in width and 73 nm in height. This specific dimensions are intentionally chosen in order for the nanoantenna not to be exactly in resonance with the excitation as resonant antennas sustain only very low incoming laser energy and are prone to a rapid degradation. Therefore, with the non-resonant nanoantenna, greater stability and reproducibility in the signal are achieved.

First, we attempt to refine the current understanding of the MPL mechanisms by measuring the nonlinearities associated to each of the autocorrelation pulses. The measurements are performed by varying the power of one pulse while the other is fixed and measuring the extra MPL, $\Delta I$, at a fixed inter pulse delay of $\Delta t$ = 0.3 ps, i.e. two pulses temporally separated. Figure 5.a shows the evolution of the extra MPL intensity, $\Delta I$, as a function of the pulse power for both configurations. We find that varying the power of the first pulse provides a power-law exponent of 3.37 while increasing the power of the second pulse features a nonlinearity close to 1.3. The asymmetry between the nonlinearities was expected according to the description of the luminescence process as a sequential multi-photon absorption[8,17,49]. However, the nonlinearity order measured at 0.3 ps interpulse delay is no longer consistent with this description. The value of 3.37 exceeds the exponent measured by single pulse excitation at 2.81 (in agreement with Figure 4.a). We ascribe the higher nonlinearity order of the first pulse to the retardation introduced by the autocorrelation measurements, which is in agreement with the report of Haug et al.[19]. In their interpretation, the order of the MPL nonlinearity originates from the intraband recombination of the hot carriers. They demonstrate an inverse proportionality between the power-law exponent and the electronic temperature of the hot carrier distribution. For single-pulse MPL measurements, electron-electron thermalization is occurring within the temporal laser pulse width where the electronic temperature is the highest. For delayed measurements, *e.g.* 0.3 ps, the relaxation of the hot distribution has already started and the electron temperature is decreasing. In qualitative agreement with Haug et al.[19], we observe in Figure 5.a an increase of the nonlinearity order. The lower power-law exponent obtained by varying the second pulse suggests that the rise of the MPL signal is more efficient because the carriers are already out of equilibrium. This overall behaviour is confirmed by Figure 5.b showing a marked increase of the power-law exponent of the process with increasing the interpulse delay $\Delta t$.

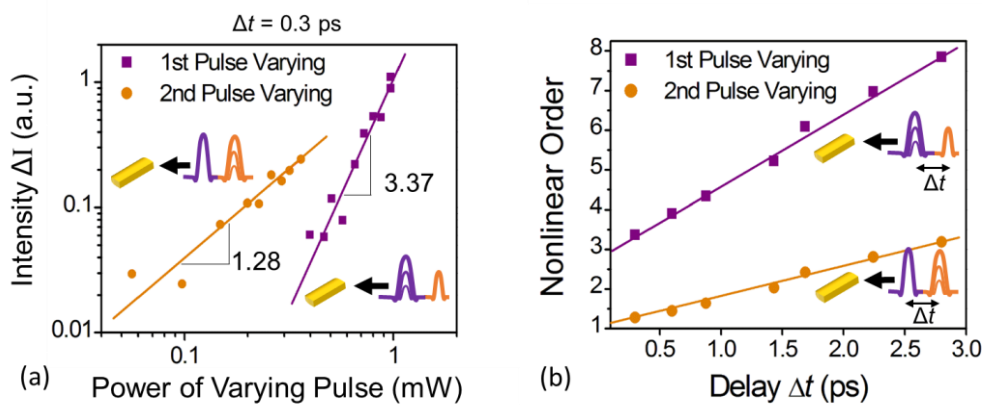

Figure 5: Nonlinearity of autocorrelation pulses. (a) Power dependence of the first pulse (squares) and the second pulse (disks) of the autocorrelation measurement. Each time a pulse power is fixed

while the other one is varying. The intensity reported on the graph corresponds to Δ*I* of the autocorrelation decaying wing, which only relates the additional MPL created by the re-excitation of hot carriers. The time delay, Δ*t*, was fixed to 0.3 ps.(b) Power-law exponents for measurements similar to that in (a) and obtained at different time delays Δ*t*. The lines represent linear fits as a guide to the eye.

Next, we set the intensity of the two pulses to the same value and record MPL spectra for different interpulse delays, Δ*t*, on the nanoantenna. In Figure 6.a, we plot the evolution of the MPL intensity (in log scale) as a function of the interpulse delay for two extreme emission wavelengths. The fact that the slopes of the curves are evolving as a function of the wavelength clearly demonstrates that the intraband relaxation dynamics of hot carriers also depends on their energies.

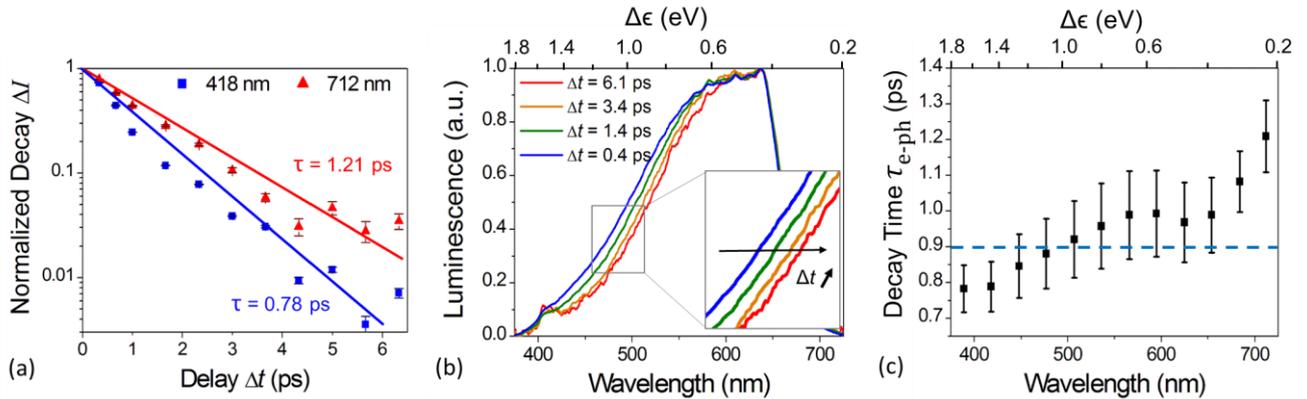

Figure 6: Coloured autocorrelation measurements. (a) Normalized decay, Δ*I*, for blue and red wavelengths, (b) nonlinear spectra for different interpulse delays, Δ*t*, and (c) calculated lifetime of hot carriers depending on their energies. The top horizontal axes indicate the lower boundary of the electron-hole pair energy: $\Delta\epsilon = hc \cdot (1/\lambda - 1/\lambda_{laser})$.

The measured spectra with different interpulse delays, Δ*t*, are presented in a normalized graph in Figure 6b. The graph unambiguously reveals a red shift for increasing delays. This is a visual and intuitive confirmation that the hot carrier distribution is getting narrower overtime. The temporal evolution of the hot-carrier temperature is accompanied with a shifting to lower energies[19], which is in agreement with an energy transfer to the phonons.

Figure 6.c, then, shows the spectral dependence of the hot carrier relaxation time, which is found to increase by 60 % when the emission energy decreases by 1.5 eV (from c.a. 400 to 700 nm). In other words, the relaxation of low-energy hot carriers is much longer than that of energetic ones. This is confirmed by the fact that hot carriers live longer in states closer to the Fermi level.[43] The substantial variation of the relaxation time can dramatically alter the efficiency of hot-carrier-assisted processes. It also implies that there is a trade-off between the energy that hot carriers can provide and the duration for which they can be used. This consideration needs to be taken into considerations in the design of hot-carrier devices and applications.

Despite the smooth trend in the decay time, $\tau_{e-ph}$, a dip is observed at around 625 nm. Striking similarities have been reported for electron-electron relaxation times, $\tau_{e-e}$, on copper and gold.[45] At the dip energy direct interband transitions are allowed whereas non-direct intraband transitions occurs otherwise. A supplementary and intuitive explanation suggests that the local dip relates to the local curvature of the band structure that the hot carriers relax on. Lastly, the local dip could stem from localized surface plasmon resonance. LSPR increases the radiation efficiency and may potentially induce faster dynamics. In additional measurements that we have performed, the LSPR scattering peak of this nanoantenna is located around 640 nm. At LSPR scattering peak, the nanoantenna is maximally coupled to photons. Therefore carriers at the LSPR energy may be converted to photons through this radiation decay channel, which is faster than carrier-phonon relaxation. Nevertheless further experimental efforts are required to rigorously unveil the origin of such a dip.

**CONCLUSIONS**

In this paper we support that the current understanding of MPL needs to take into account hot carriers-induced nonlinearities. This should open doors to further experimental and theoretical studies to refine the description of the MPL mechanisms. We also demonstrate the existence of a relationship between the energy of hot carriers and their relaxation time in monocrystalline gold nanoantennas by means of coloured (i.e. spectrally resolved) autocorrelation measurements. The present work has direct implications in hot-carrier science. It provides a proof of principle for the fabrication of a reliable and reproducible nanosources of hot carriers with predictable characteristics. Additionally, the energy-relaxation rate relationship shown here has profound repercussions. It points out the compromise between the energy that can be harvested from hot carriers and the duration of its use. Future applications will have to take this fact into account. Specific hot-carrier lifetimes may be selected by either connecting nanoantennas to quantum dots or to band-gap-engineered semiconductor structures[5]. This is a potential means of extracting carriers with a specific energy and therefore a defined relaxation time. Last, coloured autocorrelation measurements position the MPL autocorrelation technique as a valuable tool to probe hot-carrier dynamics within plasmonic nanostructures owing to their broad emission spectrum compared to linear two-colour pump-probes techniques which would require the probe to span the whole visible range.

**ACKNOWLEDGEMENTS**

This work has been supported by the Labex Action ANR-11-LABX-0001-01, the Agence Nationale de la Recherche (PLACORE ANR-13-BS10-0007), the regional funding through the PARI II Photcom and the European Research Council under the European Community's Seventh Framework Program FP7/ 2007-2013 Grant Agreement No. 306772.
The authors would like to thank Dr. F. Billard and Dr. E. Hertz for their assistance on the dispersion compensation. We would like to thank Dr F. Herbst and Dr R. Chassagnon for their assistance on EDX and TEM analyses, respectively.